\documentclass[12pt]{article}

\usepackage[utf8]{inputenc}
\usepackage{textcomp}
\usepackage{amsmath}
\usepackage{physics}
\usepackage{amssymb}
\usepackage{graphicx}
\usepackage{color}
\usepackage{dsfont}
\usepackage{bbm}
\usepackage{siunitx}
\usepackage{wrapfig}
\usepackage{notes2bib}
\usepackage[colorlinks=true,citecolor=blue,linkcolor=blue,hypertexnames=false]{hyperref}

\newcommand{\br}{{\bf r}}
\newcommand{\bk}{{\bf k}}
\newcommand{\bq}{{\bf q}}
\newcommand{\bF}{{\bf F}}
\newcommand{\bA}{{\bf A}}
\newcommand{\bB}{{\bf B}}

\newcommand{\ve}{{\varepsilon}}
\newcommand{\vp}{{\varphi}}
\def \be{\begin{equation}}
\def \ee{\end{equation}}

\usepackage{scicite}

\usepackage{times}

\newenvironment{sciabstract}{%
\begin{quote} \bf}
{\end{quote}}

\title{Direct Geometric Probe of Singularities in Band Structure}

\author
{Charles D. Brown$^{1,2,4\ast}$, Shao-Wen Chang$^{1,2}$, Malte N. Schwarz$^{1,2}$,\\ Tsz-Him Leung$^{1,2}$, Vladyslav Kozii$^{1,3}$, Alexander Avdoshkin$^1$,\\ Joel E. Moore$^{1,3}$, Dan Stamper-Kurn$^{1,2,3\ast}$\\
\\
\normalsize{$^{1}$Department of Physics, University of California, Berkeley}\\
\normalsize{366 Physics North, Berkeley, CA 94720, USA}\\
\\
\normalsize{$^{2}$Challenge Institute for Quantum Computation}\\
\normalsize{University of California, Berkeley CA 94720, USA}\\
\\
\normalsize{$^{3}$Materials Sciences Division}\\
\normalsize{Lawrence Berkeley National Laboratory, Berkeley, CA 94720, USA}\\

\\
\normalsize{$^{4}$Department of Physics, Yale University}\\
\normalsize{217 Prospect Street, New Haven, CT 06520, USA}\\
\\
\normalsize{$^\ast$To whom correspondence should be addressed;}
\\ \normalsize{E-mail:  charles.d.brown@yale.edu, dmsk@berkeley.edu.}
}

\date{}

\begin{document} 


\baselineskip24pt


\maketitle 

\newpage


\begin{sciabstract}
A quantum system’s energy landscape may have points where multiple energy surfaces are degenerate, and that exhibit singular geometry of the wavefunction manifold, with important consequences for the system’s properties. Ultracold atoms in optical lattices have been used to indirectly characterize such points in the band structure. Here, we measure the non-Abelian transformation produced by transport directly through the singularities. We accelerate atoms along a quasi-momentum trajectory that enters, turns, and then exits the singularities at linear and quadratic band touching points of a honeycomb lattice. Measurements after transport identify the topological winding numbers of these singularities to be 1 and 2, respectively. Our work introduces a distinct method for probing singularities that enables the study of non-Dirac singularities in ultracold-atom quantum simulators.
\end{sciabstract}



\section*{Main Text:}

Energy surfaces are used to describe the structure and dynamics of quantum systems whose Hamiltonians contain one or more continuous parameters.  Notable examples include band structure, which describes the motion of single particles within a crystal as a function of their quasi-momentum, and the potential energy surfaces that describe molecules as a function of their nuclear coordinates.  Each point on an energy surface corresponds to an eigenenergy and an eigenstate of the physical system.  Although the energies themselves are highly important for explaining material~\cite{xiao_berry_2010,bansil_colloquium_2016} and chemical~\cite{yarkony_conical_2001,wort04conical,domc12conical,schu18conical} properties, so too are the local geometry and global topology of the eigenstate manifolds.

The geometry of an eigenstate manifold can be revealed through transport of a quantum state along a smooth path of parameters that define the system's Hamiltonian. This transport is generally nonholonomic, meaning that the state generated by transport from an initial to a final point depends on the path along which the system was transported.  Such transport has been explored mainly in the two limiting cases in which the energy spectrum of a system is either largely gapped~\cite{simon_holonomy_1983,berry_quantal_1984} or entirely gapless~\cite{wilc84nonabelian} along a closed loop in parameter space.  In the former limit, the state-space geometry generates a Berry phase; in the latter, the nonholonomy generalizes to a Wilson loop operator describing a path-dependent rotation within the degenerate subspace. In terms of $| u^m_{\bq}  \rangle$, the cell-periodic of the Bloch wavefunction that describes a single particle, both types of dynamics derive from the Berry connection matrix, $\bA^{nm}_{\bq} \equiv i\langle u^n_{\bq}|\partial_{\bq}| u^m_{\bq}  \rangle$, which expresses the local geometry of state space. Here, focusing on the case of band structure, $n$ and $m$ are band indices, and $\mathbf{q}$ is the quasi-momentum.  In the gapped limit, the Berry phase is determined solely by one (Abelian) diagonal element of this matrix; in the gapless limit, off-diagonal elements enter, leading to non-Abelian state rotations~\cite{bohm_geometric_2003-1}.

\begin{figure} 
\centering
\includegraphics[scale=1.5]{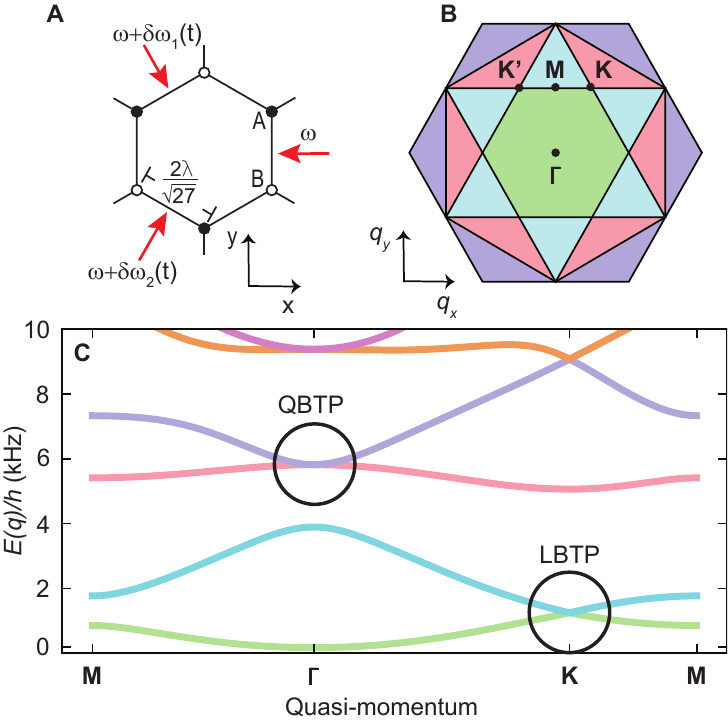}
\caption{Experimental scheme. (\textbf{A}) Illustration of an optical honeycomb lattice with two sites (A and B) in the unit cell, formed by overlapping three $\lambda=1064$ nm wavelength light beams (red arrows).  Offsetting the optical frequencies of two lattice beams by $\delta \omega_{1,2}(t)$ accelerates the lattice and drives lattice-trapped atoms through a trajectory in quasi-momentum. (\textbf{B}) The $n= \{1,2,3,4\}$ Brillouin zones of the honeycomb lattice are shown in green, blue, red and purple, respectively. (\textbf{C}) The band structure of the honeycomb lattice (plotted with  potential depth of $20~\mathrm{kHz} \times h$)  exhibits an LBTP in the $s-$orbital band manifold at $\mathbf{q}=\mathbf{K}$, and a QBTP in the $p-$orbital band manifold at $\mathbf{q}=\mathbf{\Gamma}$.}
\label{fig:scheme}
\end{figure}

In this work, we explore the nonholonomy of transport through a state space containing singular points of degeneracy.  One example of such points, which we probe experimentally, is the Dirac points of degeneracy between the $n=1$ and $n=2$ bands of the two-dimensional honeycomb lattice, lying at the $\mathbf{K}$ and $\mathbf{K}^\prime$ points of the Brillouin zone (Fig.\ \ref{fig:scheme}).  Away from these points, the energy gap between the touching bands grows linearly with quasi-momentum.  The singular state geometry around each linear band touching point (LBTP) has profound implications for the material properties of graphene, e.g.\ related to Klein tunneling of electrons through potential barriers~\cite{katz06klein} and the appearance of a half-integer quantum Hall effect~\cite{zhan05graphene}.  The Dirac point of the honeycomb lattice has been explored also in ultracold-atom experiments~\cite{tarruell_creating_2012-1}, including interferometric measurements of the Berry phase produced along trajectories that circle the Dirac point~\cite{duca_aharonov-bohm_2015-1} and direct mapping of the Bloch-state structure across the Brillouin zone~\cite{PhysRevLett.118.240403,li_bloch_2016,flaschner_experimental_2016}.

Crystalline materials may also host a singular quadratic band-touching point (QBTP), about which the energy gap between two bands grows quadratically with quasi-momentum. As before, the QBTP can profoundly affect material properties.  For example, the singular QBTP is predicted to produce an anomalous Landau level spectrum ~\cite{rhim_quantum_2020-1}.  Interactions can destabilize a QBTP, leading to topologically-protected edge states, nematic phases and both quantum anomalous Hall and spin phases~\cite{sun_time-reversal_2008,sun_topological_2009,murray_renormalization_2014,shah_renormalization_2021}. The role of QBTPs is being investigated intensely in both untwisted and twisted bilayer graphene \cite{mccann_landau-level_2006,mccann_electronic_2013,pujari_interaction-induced_2016,hejazi_landau_2019,hejazi_multiple_2019}. Despite their importance, QBTPs have remained unexplored in ultracold-atom systems.  

Here, using ultracold atoms within an optical lattice, we  demonstrate that transport of a quantum state through a singular band touching point leads to a non-Abelian, coherent state rotation between bands, with the rotation depending on the relative orientation of path tangents entering and exiting the singular point.  Further, we show that this dependence characterizes and distinguishes the Bloch-state geometry surrounding linear and quadratic band-touching singularities.

Let us exemplify our approach by considering the $s$-orbital LBTP of a two-dimensional honeycomb optical lattice  (Fig.~\ref{fig:scheme}).  To probe this Dirac point, we prepare an optically trapped $^{87}$Rb Bose-Einstein condensate and then slowly ramp up an overlain static honeycomb lattice, placing the atoms initially at the $\mathbf{\Gamma}$-point of the $n=1$ band.  Next, we apply a fictitious force to the gas by accelerating the optical lattice potential to a velocity $\mathbf{v}_\mathrm{lat}(t)$. Although the atoms remain at zero quasi-momentum in the laboratory frame, they evolve to non-zero velocity $\mathbf{v} = \hbar \mathbf{q}/m = - \mathbf{v}_{\mathrm{lat}}$ in the lattice frame, with $\mathbf{q}$ being the lattice-frame quasi-momentum and $m$ being the atomic mass.

\begin{figure} 
\centering
\includegraphics[scale=1]{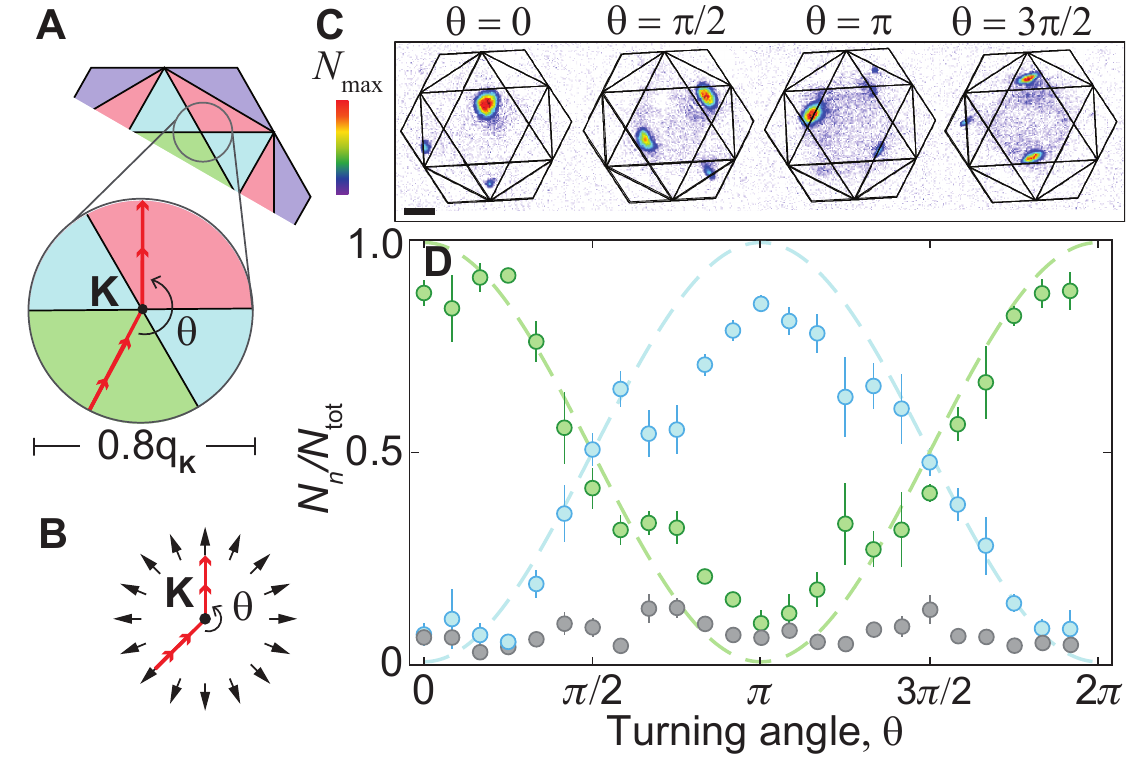}
\caption{Non-Abelian state rotations around a Dirac point. (\textbf{A}) Enlarged view of Brillouin zone map. Atoms are loaded into the state $n=1, \mathbf{q}=\mathbf{\Gamma}$ in the lattice, and transported (trajectory marked by red arrows) at constant accelerations from $\mathbf{\Gamma} \rightarrow \mathbf{K}$ and then from $\mathbf{K}$ to a final point lying on a circle of diameter $0.8\norm{\mathbf{q_{K}}}$ centered at the Dirac point. In this particular experiment, the atoms evolve between the $n=1$ and $n=2$ bands. Here, the color of the Brillouin zones does not indicate the state of the atomic wavepacket along the trajectory. Rather, the color scheme is used to interpret the band index of atoms after a band mapping measurement. (\textbf{B}) Bloch states near the Dirac point are described as pseudo-spin-1/2 states in an pseudo-magnetic field (black arrows) that points radially outward from, and wraps once around, the Dirac point. (\textbf{C}) Band mapping images at the final quasi-momentum, with overlain Brillouin zone maps, show the band population vary with turning angle $\theta$. A third, short-length and adiabatic translation step (not shown) ensures band mapping does not occur near a Brillouin zone boundary so that there is no ambiguity in the band index of atoms (see Section 3 of~\bibnotemark[SM]). The spatial scale is indicated by the black scale bar of length 0.1 mm. (\textbf{D}) Fractional band populations $N_{n}/N_{\mathrm{total}}$ ($n=1$: green, $n=2$: blue, sum of other bands: gray) vs.\ $\theta$. Means and standard mean errors determined from 7 repeated measurements. The green and blue dashed lines show a $\mathrm{cos}^2(\theta/2)$ and $\mathrm{sin}^2(\theta/2)$ dependence, respectively.}
\label{fig:fig2_v10}
\end{figure}

To demonstrate the nonholonomy generated by the LBTP, we accelerate the atoms on a trajectory that proceeds at constant acceleration from $\mathbf{\Gamma}$ to $\mathbf{K}$ (at quasi-momentum $\mathbf{q}_\mathbf{K}$), and thence at a different constant acceleration to 24 equally-spaced points on a circle that lie at distance of $0.4~\norm{\mathbf{q}_K}$ from the $\mathbf{K}$-point.  The turning angle between the rays entering and leaving the $\mathbf{K}$-point, as defined in Fig.~\ref{fig:fig2_v10}A, is varied over $\theta \in [0, 2 \pi]$.  We then perform ``band mapping" by smoothly ramping off the lattice potential at the fixed final quasi-momentum, i.e.\ with the lattice at a constant final laboratory-frame velocity.  This ramp maps the population in each band onto a distinct momentum state.  Measuring this momentum distribution quantifies band populations in the moving lattice.

Transport along paths passing through the singular LBTP leads to interband transitions that vary with the turning angle (Figs.~\ref{fig:fig2_v10}, C and D).  For trajectories that enter the singularity and then reverse onto themselves ($\theta = 0$), the population remains nearly entirely in the initial $n=1$ band.  For trajectories that continue with constant tangent through the singularity ($\theta = \pi$), the atoms undergo a near complete transition to the upper $n=2$ band (seen also in Ref.\ \cite{jotzu_experimental_2014}). Over the full range of $\theta$, each population undergoes one cycle of oscillation.

The unit-cell wavefunction of the $n=1$ and $n=2$ Bloch states near the Dirac point can be represented as a pseudo-spin-1/2 vector, with $s$-orbital Wannier states at the lattice sites A and B representing the up- and down-spin basis states.  In this basis, the Bloch states are eigenstates of the Hamiltonian $H_\mathrm{LBTP} = - \mathbf{B}(\mathbf{q}) \cdot \boldsymbol{\sigma}$ where $\mathbf{B}(\mathbf{q})$ is a pseudo-magnetic field that lies in the transverse pseudo-spin plane and ${\boldsymbol \sigma}$ is the vector of Pauli matrices. $\mathbf{B}(\mathbf{q})$ has a magnitude $B = \hbar v_{\mathrm{g}} |\mathbf{q}-\mathbf{q}_{\mathbf{K}}|$ that varies linearly with distance from the singularity, and as shown in Fig.~\ref{fig:fig2_v10}B, it has an orientation (in the proper gauge) that is radially outward from $\mathbf{K}$.  Here, $v_{\mathrm{g}}$ is the group velocity near the Dirac point.  The $2 \pi$ rotation of $\mathbf{B}(\mathbf{q})$ about the Dirac point is responsible for the $\pi$-valued Berry phase of trajectories that encircle the Dirac point \cite{duca_aharonov-bohm_2015-1}.

This pseudo-spin model explains our observations.  The atomic pseudo-spin entering the Dirac point along a ray experiences a pseudo-magnetic field whose orientation $\mathbf{n}$ remains constant and whose magnitude smoothly tunes to zero. Under this field, the initial-state pseudo-spin remains aligned along $\mathbf{n}$. Departing the Dirac point, the pseudo-spin experiences a magnetic field along a new orientation $\mathbf{m}$, with $\mathbf{n}\cdot\mathbf{m} = \cos\theta$, and a magnitude increasing linearly with time. The pseudo-spin is thus placed in a superposition of eigenstates, with population $\cos^2(\theta/2)$ in the $\mathbf{m}$-oriented pseudo-spin eigenstate ($n=1$ band) and $\sin^2(\theta/2)$ in the orthogonal state ($n=2$ band). This simple prediction is in good agreement with our data (Fig. \ref{fig:fig2_v10}D), with residual differences accounted by numerical simulations (Fig. \ref{fig:Sim1}) of the dynamics of non-interacting atoms over the finite duration of our experimental stages~\bibnote[test]{The maximum duration of our experiment was limited by the observed decay of atoms from the excited-energy Bloch states populated during lattice acceleration.}.

We find that passage through the Dirac point produces a phase-coherent superposition of band states. Such coherence is demonstrated by allowing the atoms to evolve at the final point of the trajectory for a variable time before measuring populations in a basis different from the local energy eigenbasis. Temporal oscillations in these measurements, with a frequency matching the calculated gap between the $n=1$ and $n=2$ bands, demonstrate the coherence of the atomic state following transport (Fig.~\ref{fig:Coherence}).

The energy-time uncertainty relation places a bound on how finely the singular point can be located by our method. We consider a trajectory where the acceleration has magnitude $a$ near the singularity.  The system spends a time $\delta t \sim (\hbar/ma) \delta q$ within $\delta q$ of the singularity; the energy gap has magnitude $\delta E \sim \hbar v_{\mathrm{g}} \delta q$ in that vicinity.  Setting $(\delta t) (\delta E) \sim \hbar$ establishes that the band structure is effectively gapless within a quasi-momentum distance of $\delta q =R\sim \sqrt{m a\hbar/ v_{\mathrm{g}}}$ of the singularity. That is, the nonholonomy generated by the singular point should be observed also for finite time trajectories that pass within the effective radius, $R$, of the singularity.

\begin{figure} 
\centering
\includegraphics[scale=1.5]{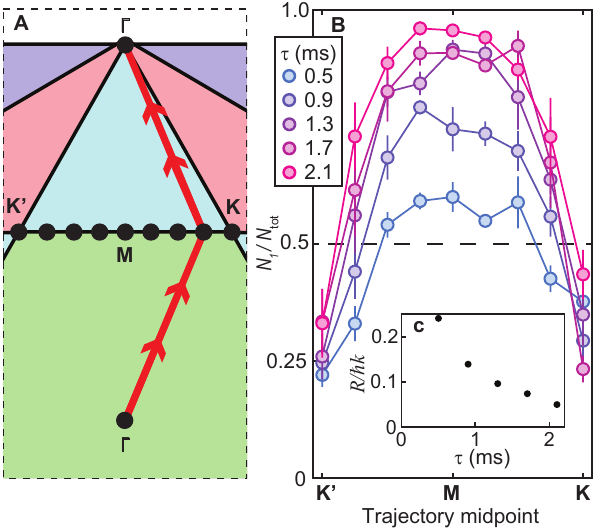}
\caption{Effective size of a Dirac singularity. (\textbf{A}) Illustration of quasi-momentum trajectories over which the atoms are transported for the measurements in \textbf{B}. Each trajectory connects two different $\mathbf{\Gamma}$-points, but traverses one of nine equally spaced points along the $\mathbf{K^\prime} - \mathbf{M} - \mathbf{K}$ line. (\textbf{B}) Fractional $n=1$ band population plotted, for different trajectory traversal times $\tau$, against the trajectory midpoints from (\textbf{A}).  Means and standard mean errors are generated from 3-5 repeated measurements. (\textbf{C}) An effective radius $R$ (plotted normalized by $\hbar k = h/\lambda$) is defined  for each $\tau$ by the distance along the $\mathbf{K^\prime} - \mathbf{M} - \mathbf{K}$ line for which the threshold $N_{1}/N_{\mathrm{total}}=0.5$ is fulfilled.  $R$ diminishes with larger $\tau$.}
\label{fig:fig3_v5}
\end{figure}

We measure $R$ by driving the atoms along a family of trajectories, shown in Fig.~\ref{fig:fig3_v5}A, that connect between the initial $\mathbf{\Gamma}$-point to a final $\mathbf{\Gamma}$-point that is one reciprocal lattice vector away, and performing band mapping measurements at the final point. These trajectories cross the boundary between the first and second Brillouin zones at nine equally-spaced points along the $\mathbf{K^\prime} - \mathbf{M} - \mathbf{K}$ line.  As shown in Fig.~\ref{fig:fig3_v5}B for various traversal times, $\tau$, we observe that trajectories that pass directly through either Dirac point yield a band population distribution that is independent of $\tau$ with $\sim 3/4$ of the atoms transferring to the upper band. In contrast, for traversal times that are longer and for paths that veer farther from the Dirac points, the transition between bands is increasingly suppressed, demonstrating that $R$ decreases with decreasing $a$ (increasing $\tau$). At first glance, there is an analogy between the experiment of Fig.~\ref{fig:fig3_v5} and Majorana losses. The study presented in Fig.~\ref{fig:fig3_v5} is quite close to Majorana’s description~\cite{majo32} and to the picture of the “Majorana hole” that appears in a spherical quadrupole magnetic trap. However, a difference between our experiment and magnetic traps is that in traps the spin flips that occur because of transport through the state geometry of the system lead to loss, whereas, in our experiment they lead to transitions between trapped bands of the lattice.

The singularity at an LBTP can be characterized by two different experimental methods: either by Berry phase measurements along trajectories that encircle the singularity~\cite{duca_aharonov-bohm_2015-1}, or, as shown here, through state rotations produced along trajectories that pass through the singularity. These two methods are related, but nonequivalent. Berry phase measurements measure the integrated Berry flux, which is determined from a diagonal element of the Berry connection matrix $\bA^{nn}_{\bq}$. A $\pi$-valued flux is found pinned to the singular point. In contrast, the non-Abelian state rotations detected in our method derive directly from the off-diagonal elements $\bA^{nm}_{\bq}$ with $n = 1$ and $m = 2$ being the two crossing bands~\bibnote[SM]{See supplementary materials on Science Online}. Further, different from Berry phase measurements, our method can be regarded as measuring the Hilbert-Schmidt quantum distance $d^2(\bq, \bq')=1 - |\langle u_{\bq'}^1 |   u_{\bq}^1 \rangle|^2$~\cite{provost_riemannian_1980,hwang_wave-function_2021} with $\mathbf{q}$ identified as a point along the input path into -- and $\mathbf{q}^{\prime}$ as a point along the exit path from -- the singularity. The oscillation of the $n = 1$ band population as a function of the turning angle reveals the quantum distance to undergo one complete oscillation between zero and unity on a contour encircling the LBTP.

The distinction between these two methods is dramatic in the case of a QBTP. Like the LBTP, a singular QBTP also carries concentrated Berry flux that is restricted, assuming time-reversal and $C_6$ symmetry, to be 0 or $\pm 2 \pi$~\cite{sun_topological_2009}. However, these values of Berry phase are undetectable via interference measurements. In contrast, our method uncovers the characteristic nonholonomy of the singular QBTP and the concomitant modulation of the quantum  distance around the singular point.

\begin{figure} 
\centering
\includegraphics[scale=1]{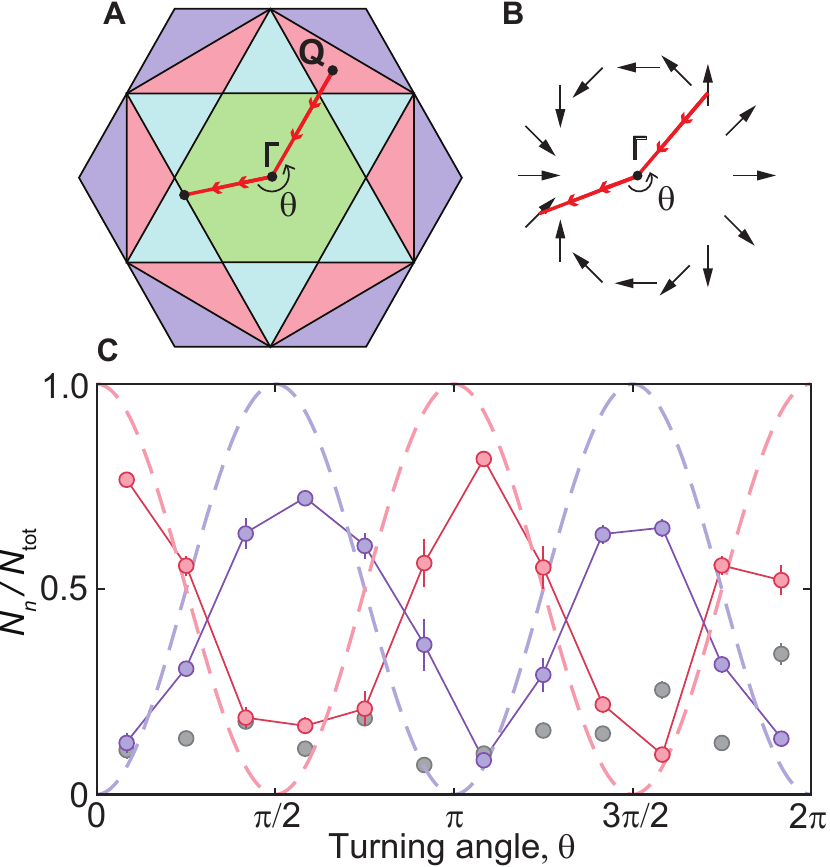}
\caption{Non-Abelian state rotations around a QBTP. (\textbf{A}) Atoms are prepared in the $n=3$ band at $\mathbf{Q}$, transported (along red arrows) to the QBTP at $\boldsymbol{\Gamma}$, and then transported to a final quasi-momentum for band mapping. In this particular experiment, the atoms evolve between the $n=3$ and $n=4$ bands. Here, the color of the Brillouin zones does not indicate the state of the atomic wavepacket along the trajectory. Rather, the color scheme is used to interpret the band index of atoms after a band mapping measurement. (\textbf{B}) The pseudo-magnetic field (black arrows) describing the Bloch state geometry wraps twice in orientation  for one revolution around the QBTP. (\textbf{C}) A plot of normalized band population as a function of $\theta$, collected by analyzing band mapping data (Fig.~\ref{fig:BM}). Red circles: $n=3$ band; purple circles: $n=4$ band; gray circles: bands with $n\neq3,4$.  Means and standard mean errors are determined from 7 repeated measurements. Dashed red (purple) lines are predictions based on a simple pseudo-spin model for $n=3$ ($n=4$) populations. Our numerical simulations suggest that the data does not reach unity oscillation amplitude owing to nonadiabaticity in the band mapping procedure (Fig.~\ref{fig:Sim1}).}
\label{fig:fig4_v6}
\end{figure}

For a singular QBTP, the geometric structure of Bloch states of the two intersecting bands at the vicinity of the singularity can again be described as those of a pseudo-spin in a pseudo-magnetic field. Different from the LBTP, here, the pseudo-magnetic field, lying in the transverse pseudo-spin plane, has a magnitude that increases quadratically with distance to the singularity, and has an orientation that wraps by an angle of $4 \pi$ along a path encircling the singularity; see Fig.~\ref{fig:fig4_v6}B.

We probe this geometric structure at the QBTP that occurs at  $\mathbf{\Gamma}$ between the $n=3$ and $n=4$ bands of the honeycomb lattice.  For this, we first load the Bose-Einstein condensate into the $n=3$ band of the lattice by ``inverse band mapping.''  In previous work \cite{leung_interaction-enhanced_2020}, such loading into excited bands was realized with \emph{moving} atoms in a \emph{static} lattice; here, we realize similar state-preparation with \emph{static} atoms loaded into a \emph{moving} lattice.  Specifically, we gradually increase the depth of a honeycomb lattice moving with velocity $\mathbf{v}_\mathbf{Q}$ that, as shown in Fig.~\ref{fig:fig4_v6}A, is located within the third Brillouin zone in the extended zone scheme. Thence, we accelerate atoms (in the lattice frame) at constant acceleration along the path $\mathbf{Q} \rightarrow \mathbf{\Gamma}$, into the QBTP, and then, turning by an angle $\theta$, accelerate the atoms at a different constant acceleration out of the QBTP and to the edge of the Brillouin zone.  Final points are chosen as ones where populations in the $n=3$ and $n=4$ bands are easily distinguished by band mapping.

We observe a nonholonomy at the QBTP that is distinct from that observed at the LBTP.  Specifically, we observe \textit{two} cycles of oscillation in the final band populations over the interval $\theta \in [0, 2\pi]$. This behavior is explained well by the pseudo-spin representation of the singular QBTP. Different from the LBTP, here, pseudo-magnetic-field orientations along the incoming ($\mathbf{n}$) and outgoing ($\mathbf{m}$) paths are related as $\mathbf{n} \cdot \mathbf{m} = \cos 2 \theta$. The nonholonomy of the QBTP now produces populations of $\cos^2(\theta)$ and $\sin^2(\theta)$ in the $n=3$ (initial) and $n=4$ bands, respectively.

The amplitude of the observed oscillation is lower than suggested by this simple theory; again, we ascribe this difference to dynamical effects of our finite-duration acceleration and band-mapping stages.  Nevertheless, the periodicity of the oscillations, combined with the known time-reversal and  $C_6$ symmetry of our lattice, unambiguously determines the topological winding number around the QBTP to be well defined and equal to 2~\bibnotemark[SM].

In conclusion, we have demonstrated transport of a quantum system through singular band-touching points with different topological winding numbers. We observe non-Abelian, coherent state rotation between bands.  The dependence of this rotation on the relative orientation of path tangents entering and exiting the singular point unambiguously measures the winding number.

Our method of probing band structure could be applied to gain insight on other band structure singularities and on interaction effects. It would be interesting to study higher-order singular band-touching points, i.e.,  between more than two bands. We deliberately minimize interaction effects in these experiments, but in future work it will be interesting to observe potential interaction-induced instabilities of Dirac points, QBTPs or other band touching points.  The path-dependent non-Abelian nonholonomy observed here may also pertain to chemical systems, where potential energy surfaces are endowed similarly with conical intersections \cite{yarkony_conical_2001,wort04conical,domc12conical,schu18conical}, suggesting a potential route for quantum state control in optically driven molecules.

\bibliographystyle{Science}
\bibliography{allrefs_CB,allrefs_x2,morebibs}

\begin{thebibliography}{10}
\bibitem{test}
The maximum duration of our experiment was limited by the observed decay of
  atoms from the excited-energy Bloch states populated during lattice
  acceleration.
\end{thebibliography}

\section*{Acknowledgments}
We thank E. Altman, M. Zaletel, N. Read and J. Harris for early insights into this work. \textbf{Funding:} We acknowledge support from the NSF QLCI program through grant number OMA-2016245 and also NSF grant PHY-1806362, and from the ARO through the MURI program (grant number W911NF-17-1-0323). C. D. B. acknowledges support from the National Academies of Science, Engineering, and Medicine Ford Postdoctoral Fellowship program.  V. K. and J. E. M. were supported by the Quantum Materials program at LBNL, funded by the U.S. Department of Energy under contract number DE-AC02-05CH11231. A. A. and J. E. M. acknowledge support from the NSF under grant number
DMR-1918065 and from a Kavli ENSI fellowship. J. E. M. acknowledges support from a Simons Investigatorship. \textbf{Author contributions:} All authors contributed substantially to the work presented in this manuscript. C.D.B., S-W. C., M.N.S. and T-H. L. acquired the data and maintained the experimental apparatus. C.D.B., S-W. C., and M. N. S. analyzed the data. C.D.B. prepared the manuscript. M. N. S., A. A. and V. K. performed the numerical calculations. A. A. and V. K. performed supporting theory calculations. J.E.M. and D. S-K. supervised the study. All authors worked on the interpretation of the data and contributed to the final manuscript. \textbf{Competing interests:} The authors declare no competing interests. \textbf{Data and materials availability:} The data and the codes used for both data analysis and theory curve generation are available at Zenodo~\bibnote{C. D. Brown, S.-W. Chang, M. N. Schwarz, T.-H. Leung, V. Kozii, A. Avdoshkin, J. E. Moore, D. Stamper-Kurn v1.1.0, Zenodo (2022); https://doi.org/10.5281/zenodo.6788172}.

\section*{List of Supplementary Materials}
Materials and Methods\\
Supplementary Text\\
Figs S1-S7\\
References \cite{browaeys_transport_2005, zhang_honeycomb_2014,sun_topological_2009, chiu_classification_2014,chiu_classification_2016,montambaux_winding_2018,solt11hexagonal}

\newpage


\setcounter{equation}{0}
\setcounter{figure}{0}
\setcounter{table}{0}
\setcounter{section}{0}

\renewcommand{\theequation}{S\arabic{equation}}
\renewcommand{\thefigure}{S\arabic{figure}}
\renewcommand{\thetable}{S\Roman{table}}

\section*{Materials and Methods}
\label{section:methods}
We prepare Bose-Einstein condensates of $1-2\times10^{4}$ $^{87}$Rb atoms in the the hyperfine state $\ket{F=1, m_F=-1}$ in an optical dipole trap with trap frequencies $\omega_{x,y,z} \approx 2 \pi \times \{23,41, 46\} $ Hz. The chemical potential of the harmonically trapped gas is roughly $0.4 \, \mbox{kHz}\times h$. We load a condensate into a honeycomb lattice with lattice depth $\approx 5E_{\mathrm{rec}}-10E_{\mathrm{rec}}$, where the photon recoil energy $E_{\mathrm{rec}}/h=1/2\pi(\hbar/2m)*(2\pi/\lambda)^2=$ 2 kHz. The lattice is formed by the mutual interference of three $\lambda=1,064$ nm wavelength laser beams propagating in the horizontal plane, intersecting at equal angles of $120^\circ$, and polarized in-plane. In ref.~\cite{solt11hexagonal} the authors describe an optical honeycomb lattice in which atoms experience a vector a.c. Stark shift that varies across the two sites of the unit cell, causing an A-B site asymmetry in potential depth. This Stark shift is also present in our system, causing a small gap calculated to be 33 Hz at the $\textbf{K}$-point. Yet, we do not observe, nor would we expect to observe such a small gap having a noticeable effect on our transport experiment. In Fig.~3 of the main text, we show that the effects of the Dirac point are seen even for trajectories that do not hit the Dirac point, and that pass entirely through gapped regions of the band structure, so long as the trajectory is traversed quickly with respect to the residual gaps near the Dirac point. We expect that trajectories that probe the Dirac point on timescales that are short compared to the inverse gap time produced by the vector a.c. Stark shift will not be affected by that gap.

We operate our lattice in a ``lattice of tubes'' configuration, with only weak confinement in the vertical direction. Our calculations suggest that, with the lattice turned on, there are approximately 90 atoms per lattice site at the center of the trap. One lattice beam has a fixed angular frequency $\omega$, while the other two beams have time-dependent angular frequencies $\omega_1(t)=\omega+\delta \omega_1(t)$ and $\omega_2(t)=\omega+\delta \omega_2(t)$. Under a closed feedback loop that dynamically varies the phase difference between the intersecting beams that form the lattice, time-dependent relative detuning between lattice beams allows us to accelerate the lattice potential reproducibly in two dimensions. The lattice acceleration drives the atoms through a two-dimensional trajectory of quasi-momentum in the lattice frame.  After transporting the atoms along this trajectory, we adiabatically ramp down the lattice potential to map Bloch states onto a plane wave basis (band mapping) and then measure the atomic velocity distribution using velocity-space focusing and resonant imaging.  This distribution reveals the lattice band populations prior to band mapping.

\section*{Supplementary Text}
In this supplementary material, we provide additional details of our experiment and the theoretical framework used to understand our measurements. In Section \ref{section:trans}, we give a detailed description of the working principle of our lattice translation setup. The time sequence that controls lattice translations is then discussed in Section \ref{section:Expseq}. In Section  \ref{section:coh_evo}, we provide evidence that the system undergoes a coherent evolution after passing a singularity. In Section \ref{section:fig3_complement}, we provide a complementary way to understand the results presented in Fig.~3 of the main text. In Section  \ref{section:QBTP_extent}, we characterize the effective size of the QBTP between the third and fourth bands, with experiments that are analogous to those carried out in Fig.~3 of the main text. In Section \ref{section:Hsim}, we provide a method for numerically simulating the time evolution of our system, which is useful for choosing a set of parameters for experiments. Finally, to shed light on the connection between our experimental results and the topological properties of the system, we provide two-band models in Section  \ref{section:2BandModel} for both the LBTP and the QBTP that reproduces the non-Abelian state rotations that we measure.

\newpage
\tableofcontents

\section{Arbitrary lattice translation}
\label{section:trans}
The optical honeycomb lattice potential is created by interfering three laser beams, with wavelength $\lambda = 1,064$ nm, that intersect at relative angles of $120^{\circ}$. Written explicitly, the potential is given by 
\be
    \begin{aligned}
        V = \left|\sqrt{\frac{2 V_{\mathrm{lat}}}{9}} \sum_{i = 1, 2, 3} \mathbf{\epsilon}_i e^{\mathrm{i} \mathbf{k}_i \cdot \mathbf{r}} \right|^2 = \frac{2}{9} V_{\mathrm{lat}} \left(3 - \sum_{i = 1, 2, 3} \cos{(\mathbf{G}_i \cdot \mathbf{r})}\right),
    \end{aligned}
\ee
where $\mathbf{k}_n = k (\cos{\theta_i}, \sin{\theta_i}, 0)$ \footnote{Here we use a Cartesian basis with $\mathbf{x}$ and $\mathbf{y}$ being unit vectors spanning the horizontal lattice plane, and $\mathbf{z}$ oriented vertically.}
are the wavevectors of lattice beams with $k = 2 \pi / \lambda, \, \theta_i = (-1 + \frac{2 i}{3}) \pi$, $\mathbf{\epsilon}_i = (\sin{\theta_i}, - \cos{\theta_i}, 0)$ are the (in-plane) polarization vectors, $V_{\mathrm{lat}}$ is the depth of the lattice, and $\mathbf{G}_i = \sum_{j,k} \epsilon_{ijk} \mathbf{k}_j$ are the reciprocal lattice vectors ($\epsilon_{ijk} $ is the antisymmetric Levi-Civita tensor). Here, the script index $i$ is not to be confused with the imaginary number $\mathrm{i}$. When one of the beams with angular frequency $\omega_1=\omega +\delta \omega_1$, is detuned by $\delta \omega_1$ relative to the beam with angular frequency $\omega$, the lattice will translate at a velocity $\mathbf{v}_1 = a \Delta \omega_1 \hat{k}_1$ in the lab frame, where $a = \frac{2 \lambda}{3}$ is the lattice constant. Here, we ignore the small change in the wavevector $\mathbf{k}_1$ due to the detuning. Similarly, detuning beam 2 by $\Delta \omega_2$ translates the lattice at a velocity of $\mathbf{v}_2 = a \Delta \omega_2 \hat{k}_2$. Therefore, by choosing an appropriate linear combination of $\mathbf{v}_1$ and $\mathbf{v}_2$, we can translate the lattice at any velocity in the lattice plane. Acceleration along a linear trajectory in quasi-momentum space corresponds to varying the detunings of the two lattice beams linearly in time.

\begin{figure}
    \centering
    \includegraphics[width=\columnwidth]{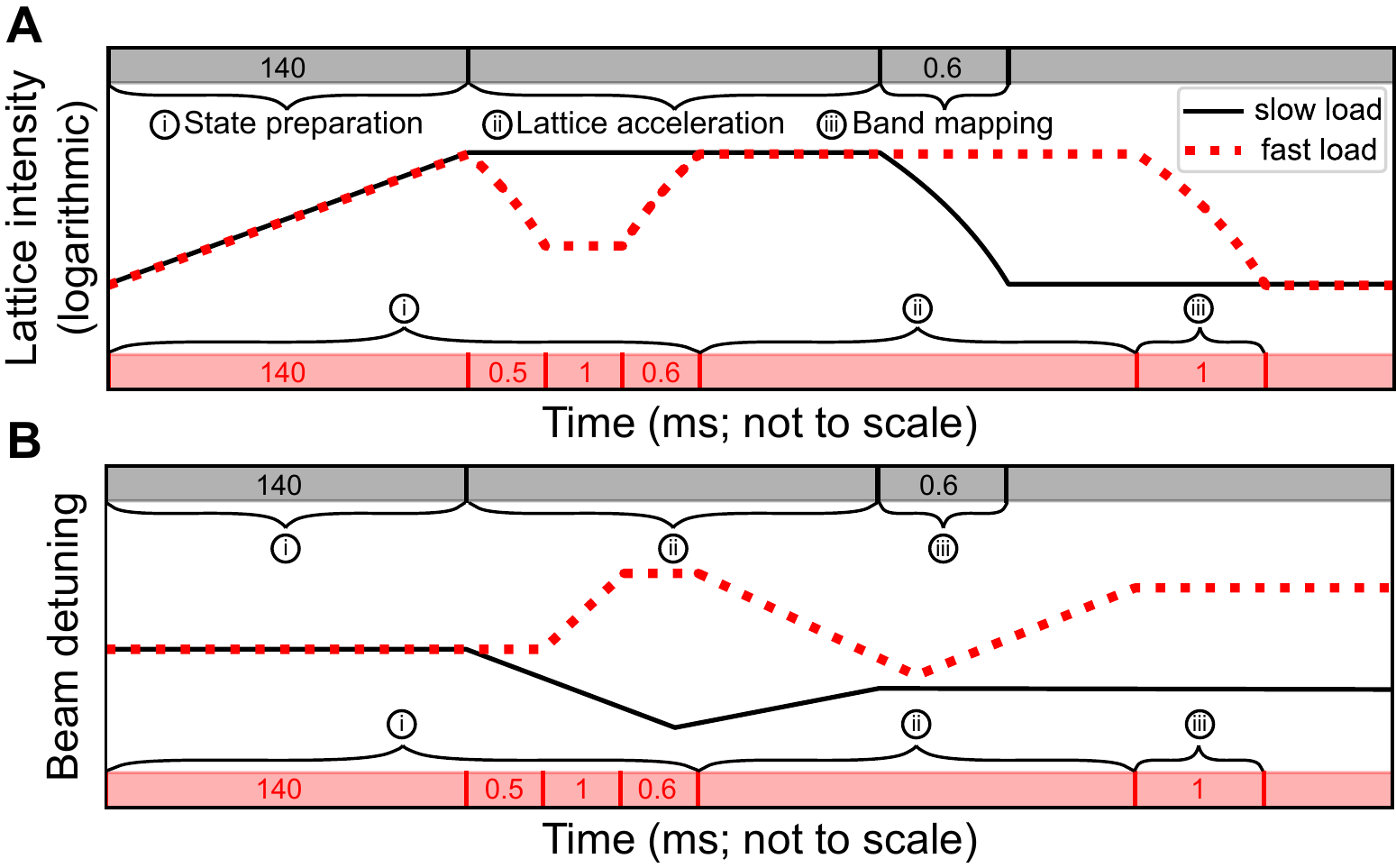}
    \caption{Experimental sequence. (\textbf{A}) Typical ramp shape of lattice beam intensity. The same ramp function is shared for all three lattice beams. (\textbf{B}) Typical ramp shape of lattice beam detuning. Generally, the detunings between two beam pairs are different, depending on the trajectory in momentum space. The shaded bars on top and bottom of both \textbf{A} and \textbf{B} indicate the time intervals (in units of milliseconds) used for each stage of the experiment: top (gray) = slow load, bottom (red) = fast load.}
    \label{fig:Exp_seq}
\end{figure}

In the experimental setup, each lattice beam is controlled by an acousto-optic modulator (AOM), which we use to control the detuning between lattice beams. We change the drive frequency of AOMs by tuning the set voltage of the voltage-controlled oscillators (VCOs) that control the AOMs, as shown in Fig.~\ref{fig:Exp_seq}B.  For each line segment in momentum space, we change the detuning linearly, which creates a uniform acceleration of the lattice. In the reference frame where the lattice is stationary, the atoms are thus accelerated in the opposite direction.

\section{Lattice translation sequence}
\label{section:Expseq}
The ramp shapes of beam intensities are given in Fig.~\ref{fig:Exp_seq}A. After creating a Bose-Einstein condensate, we first adiabatically ramp up the lattice intensity, such that all atoms are loaded into the ground band at $\mathbf{\Gamma}$; see ``slow load'' in Fig.~\ref{fig:Exp_seq}. This also allows the atoms to follow the minimum of the combined trap potential created by the optical dipole trap (ODT) beams and the lattice beams, which varies as the lattice intensity is ramped up, due to possible misalignment between ODT beams and lattice beams.

For the QBTP experiment we prepare atoms in the $n=3$ band, which can be done by rapidly loading atoms into a running lattice. In reality, to make sure that the atoms are at the minimum of the trap potential at the start of the experiment, we first load atoms into the ground band with the slow loading procedure. Next, we perform band mapping to recover a stationary BEC, before rapidly ramping up the lattice beam intensities again but with a running lattice. This is shown in Fig.~\ref{fig:Exp_seq} as ``fast load.'' With this technique, we can load into the $n = 3$ band with $> 92\%$ fidelity on average. 

\begin{figure}
    \centering
    \includegraphics[width=\columnwidth]{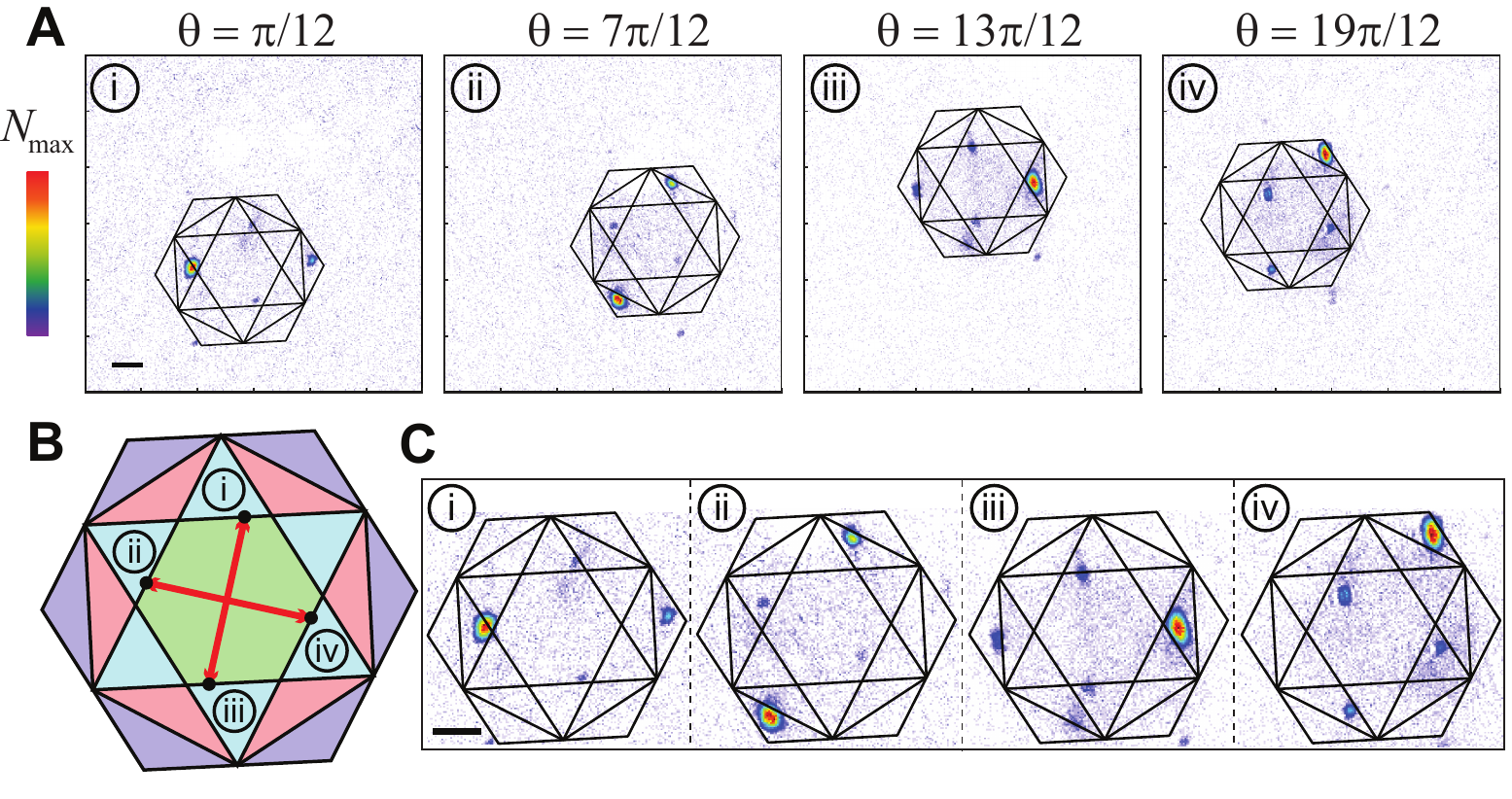}
    \caption{Band mapping data. Shown in the figure are individual measurements from our characterization of the honeycomb-lattice QBTP, as summarized in Fig.~4 of the main text. Images $\mathrm{i}$ through $\mathrm{iv}$ (in \textbf{A} and \textbf{C}) show the results of band mapping after trajectories with four different turning angles $\theta$, at the final quasi-momenta marked by black points on the Brillouin zone map in \textbf{B}. \textbf{(A)} Band mapping images obtained in the lab frame. These are absorption images of the atomic spatial distribution taken after (1) adiabatically ramping off the lattice beams (and also optical traps) at the final lattice velocity, and (2) allowing the atoms to expand in a weak harmonic magnetic confinement so as to map velocity onto position. The spatial scale of the images in \textbf{A} and \textbf{C} is indicated by the black scale bar of length 0.1 mm. The scale of the images in terms of velocity distributions is indicated by the overlain Brillouin zone maps. The color bar applies to both \textbf{A} and \textbf{C}. (\textbf{C}) The band mapping images in the lattice frame. Each of them are obtained by translating the lab-frame image in \textbf{A} by the corresponding quasi-momentum (one of the red arrows) in \textbf{B}.}
    \label{fig:BM}
\end{figure}

Band population detection (or ``band mapping'') is achieved by ramping down the lattice beams slowly enough, such that atoms in different bands are adiabatically mapped to plane waves with momenta separated by reciprocal lattice vectors. In the resulting absorption images, the positions of atom peaks then correspond to their momentum in the lab frame. In the main text, we show our band mapping measurement results in the lattice frame, which is obtained by translating the lab frame image by a distance determined by the final quasi-momentum in a lattice acceleration sequence. The positions of atom peaks in the images now correspond to their momenta in the lattice frame, and the association between band label $n$ and these peaks can then be determined by overlapping the Brillouin zone of the lattice with the image. Fig.~\ref{fig:BM} provides an example of images obtained from band mapping measurements.

In the measurement of Fig.~2C of the main text, if the angle of the outgoing trajectory would lie close to a Brillouin zone edge, we accelerate the atoms further before band mapping to a point far away from any Brillouin zone edges. This further acceleration is slow and adiabatic, such that the measured population in each band is not affected by the additional translation.

\section{Coherent evolution after population transfer at a singularity}
\label{section:coh_evo}
\begin{figure}
    \centering
    \includegraphics[width=\columnwidth]{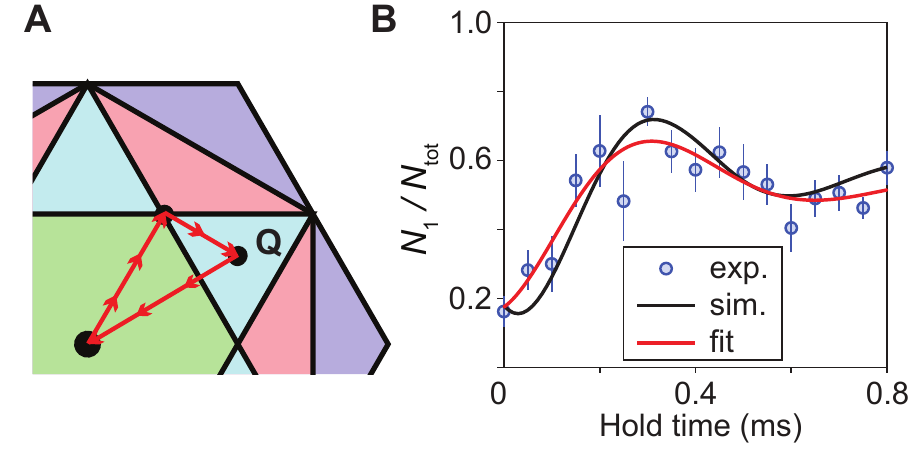}
    \caption{Detecting coherence of superposition produced by transport through the LBTP singularity. (\textbf{A}), Atoms are transported on a trajectory with three constant accelerations: $\boldsymbol{\Gamma} \rightarrow \mathbf{K}$, $\mathbf{K} \rightarrow \mathbf{Q}$, and $\mathbf{Q} \rightarrow \boldsymbol{\Gamma}$. The first two acceleration steps are done adiabatically, in 0.3 ms each. Atoms are then held at $\mathbf{Q}$ for a variable hold time. The last acceleration step is done quickly in 0.2 ms. This faster acceleration causes the atomic state (a superposition in the $n=1$ and $n=2$ bands at $\mathbf{Q}$) to be projected onto the band eigenbasis at $\boldsymbol{\Gamma}$ prior to band mapping. (\textbf{B}) $n=1$ band population seen from band mapping at $\boldsymbol{\Gamma}$. The dynamical phase acquired by the superposition state $\mathbf{Q}$ leads to temporal oscillation in the band-mapping $n=1$ population. We fit the data to a sinusoid with a decaying exponential envelope (red curve), which finds that the frequency is $1.47 \pm 0.53$ kHz, and the decay constant $319 \pm 205 \ \mathrm{\mu}$s. This frequency is in agreement with the band gap between $n = 1$ and $n = 2$ bands at $\mathbf{Q}$, which is calculated as $1.77$ kHz at a lattice depth of $V_\mathrm{lat} = h \times 24$ kHz $= 12 E_{\mathrm{rec}}$ for non-interacting atoms. The black curve is the simulation result obtained by a numerical simulation of the entire transport, hold time, and projective measurement process, in the non-interacting limit, as described in Sec.~\ref{section:Hsim}. The simulation does not account for decay: To make a more direct comparison between theory and experiment, we apply exponential damping to the simulation result using the decay constant found from our fit to experimental data.}
    \label{fig:Coherence}
\end{figure}

Here, we show that after the population transfer at $\mathbf{K}$, the atoms in bands $n=1$ and $n=2$ still interfere coherently. The trajectory used for the coherence test is shown in Fig.~\ref{fig:Coherence}A. After creating an even superposition in the lowest two bands by turning at $\mathbf{K}$, the atoms are accelerated to the center of the second Brillouin zone, at point $\mathbf{Q}$ in momentum space, and held there for a varying amount of time $t$. During the hold, the state acquires a relative phase $\varphi = t\Delta \omega_\mathbf{Q}$ between the $\ket{n = 1}$ and $\ket{n = 2}$ components, where $\hbar \Delta \omega_\mathbf{Q}$ is the band gap between the lowest two bands at  $\mathbf{Q}$. Then, we accelerate the atoms quickly back to $\mathbf{\Gamma}$, with an acceleration large enough such that the state is projected onto the basis states formed by the Bloch states at $\mathbf{\Gamma}$. Finally, we perform a band mapping measurement to project the phase difference into a population difference between bands $n=1$ and $n=2$. In Fig.~\ref{fig:Coherence}B, we indeed see an oscillation in the measured population that is consistent (with the empirically added decay) with what is expected from theory. 

\section{Complementary representation of data in Fig.~3 of the main text}
\label{section:fig3_complement}
\begin{figure} [h]
    \centering
    \includegraphics[width=\columnwidth]{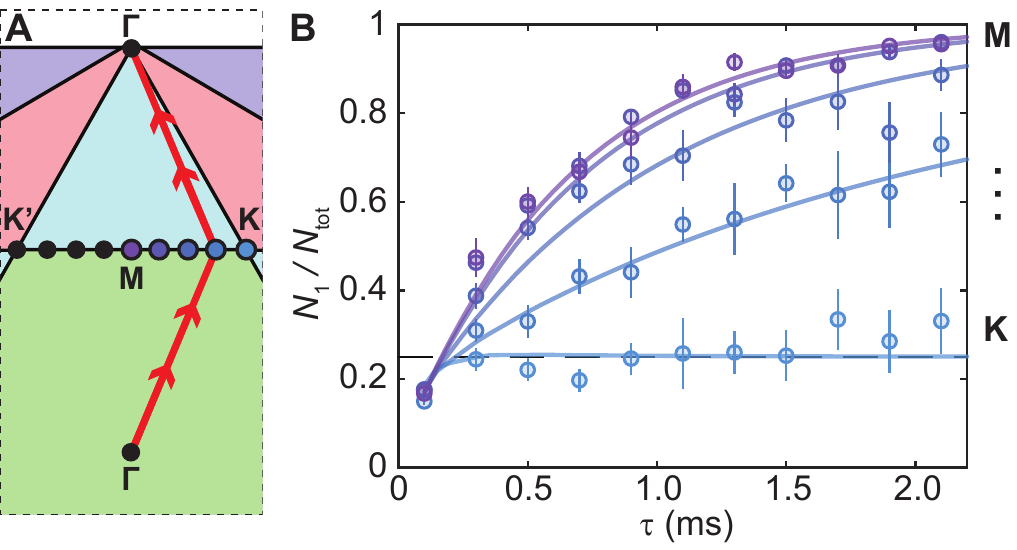}
    \caption{Complementary representation of the same data set of Fig.~3 in the main text. (\textbf{A}) Trajectories taken in the experiment copied here for reference. (\textbf{B}) normalized $n=1$ band population as a function of acceleration time $\tau$. Only data for trajectories that go through points between $\mathbf{K}$ and $\mathbf{M}$ are shown in the plot, for simplicity. Here, the solid lines are the simulation results obtained by numerical simulations of the entire transport, hold time, and projective measurement process, in the non-interacting limit, as described in Sec.~\ref{section:Hsim}. For the shortest acceleration time used, we expect the $n=1$ band population to deviate from 0.25 (dashed line), predicted by the two-level model in Sec.~\ref{section:2BandModel}, because of excitation to higher bands.}
    \label{fig:all_deltas}
\end{figure}

In Fig.~3 of the main text, we show a measurement of the effective size of a Dirac point singularity by plotting the population in the lowest band against the midpoint of various trajectories (the trajectories are shown both in Fig.~3 of the main text, and in Fig.~\ref{fig:all_deltas}A). Instead, here we provide a different visualization of the data. In Fig.~\ref{fig:all_deltas}B we plot the $n=1$ band population normalized by the total atom number against the acceleration time used to traverse the trajectories in Fig.~\ref{fig:all_deltas}A. A projection of the initial state $\ket{\psi^1_\mathbf{\Gamma}}$ onto the $n=1$ Bloch state at the final $\mathbf{\Gamma}$-point in the two-band model gives 0.25, which is the expected normalized population in the $n=1$ band under the sudden approximation. For trajectories far away from singularities, the adiabatic theorem predicts that the $n=1$ band population should increase with longer acceleration times, which we indeed observe in our experiments. On the other hand, when we cross $\mathbf{K}$ in the trajectory, the band gap between $n = 1$ and $n = 2$ bands vanishes, and the adiabatic theorem no longer applies.

\section{Momentum space extent of the QBTP}
\label{section:QBTP_extent}
\begin{figure}
    \centering
    \includegraphics[width=\columnwidth]{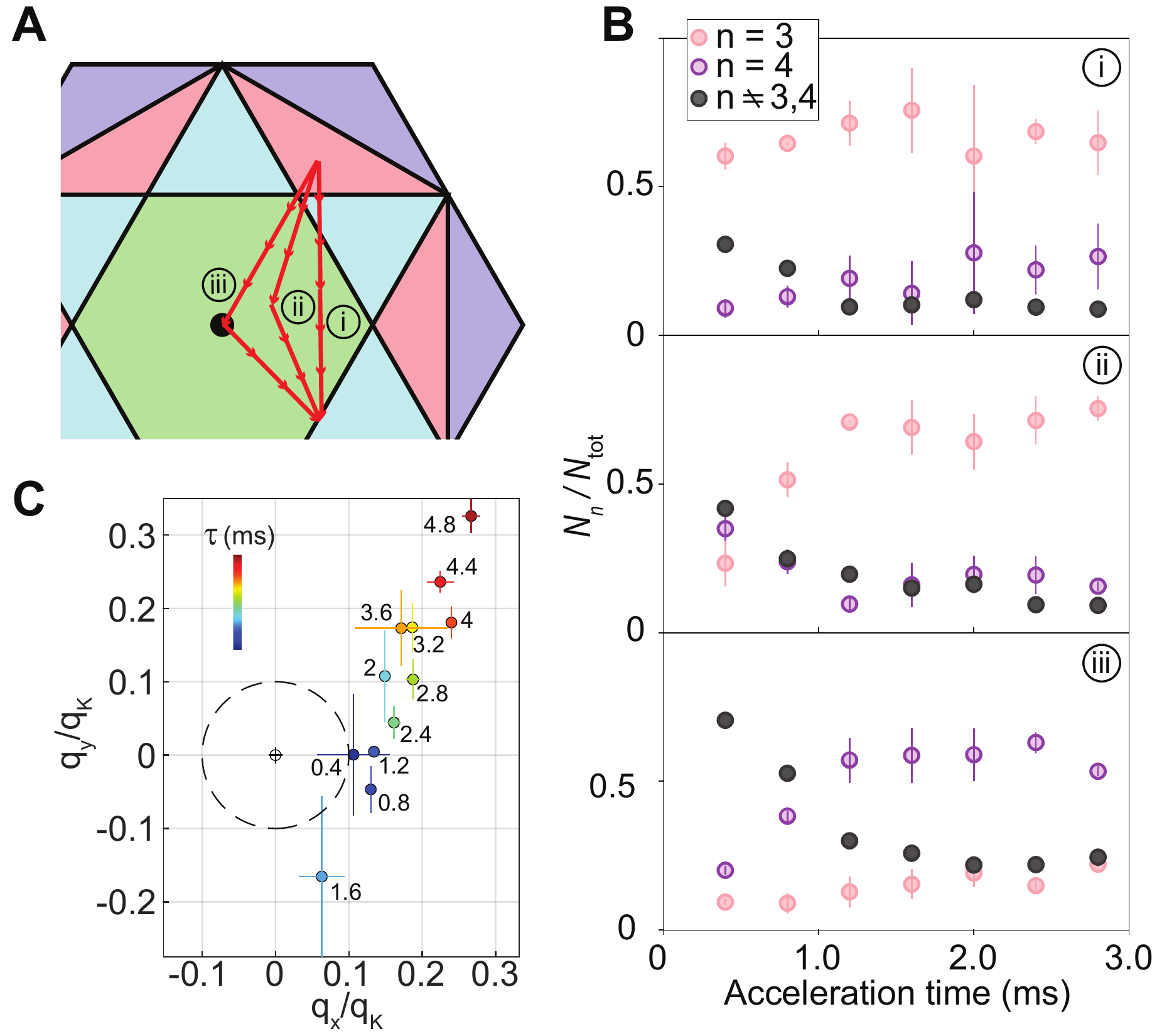}
    \caption{Momentum-space extent of the QBTP. (\textbf{A}) Trajectories chosen for the experiments. (\textbf{B}) The population in each band as a function of acceleration time $\tau$. Here, $\tau$ is the total time taken to accelerate the atoms from the initial state to the final state. For this choice of band mapping point, only the third and fourth bands are uniquely distinguished. Error bars are one standard error. (\textbf{C}), Acceleration in lab frame. As described in Sec.~\ref{section:Expseq}, the positions of atom peaks in images correspond to their lab frame velocities. Therefore, by tracing the position of one of the peaks in band mapping images in the lab frame, we can measure the change in lab-frame momentum that atoms obtain during the experiment. The figure is plotted with data from \textbf{B}-i, and different colors are assigned for each acceleration time labeled next to the data points. The radius of the dashed circle is $0.1 \|\mathbf{q_K}\|$, centered at the position corresponding to zero momentum in lab frame, which is determined by repeated measurements of the atomic wavepacket's position without lattice translation in a separate set of experiments.}
    \label{fig:QBTP_delta}
\end{figure}

To characterize the singular behavior of the QBTP, we perform a measurement similar to the one shown in Fig.~3 of the main text. As seen in Fig.~\ref{fig:QBTP_delta}B, for the same acceleration time $\tau$, more atoms stay in the third band for the trajectories that are farther away from $\mathbf{\Gamma}$, since the condition for adiabatic evolution is better satisfied. At very short acceleration times, transitions to higher bands are significant, which is also captured by simulations. At longer acceleration times ($\tau > 1.4$ ms), the actual quasi-momentum of atoms in the lattice deviates from the set values due to acceleration in the lab frame, which appears in the images as an overall shift of peak positions after band mapping (Fig.~\ref{fig:QBTP_delta}C). There are two immediate consequences: first, the actual trajectory is different from expected; second, band mapping becomes unreliable if the actual final quasi-momentum is close to the boundary between the third and fourth Brillouin zones. Therefore, we only plot the measurement results up to $\tau = 2.8$ ms, after which we have difficulties assigning band indices to different wave packets in the images. We attribute this acceleration to the fact that atoms are dragged along the direction of lattice translation during the sequence (plus corrections due to group velocity of atoms in the lattice)~\cite{browaeys_transport_2005}. The dragging effect grows with the duration of the experiment. The QBTP experimental sequence is significantly longer than the LBTP experiment (there are more steps in the procedure and the band mapping time is necessarily longer), so the dragging effect is more pronounced. Once the atoms are dragged away from the potential minimum of the optical trap, this confining potential applies an acceleration to the atoms. Finally, we indeed observe that, like the LBTP, the region in which band population rotations occur is confined at the QBTP. 

\section{Simulations via full Hamiltonian construction}
\label{section:Hsim}
The dynamics of the system is governed by the time dependent Schrödinger-equation:
\be
    \mathrm{i} \hbar \partial_t \ket{\psi} = H(t) \ket{\psi}.
\ee

We divide the time sequence into small time steps with effectively constant Hamiltonian $H(t)$ such that we can use the time-independent Schrödinger equation. Bloch's theorem allows us to write the Hamiltonian in the plane wave basis $\braket{\mathbf{r} | \mathbf{G}_{(n,m)}} = \exp{(\mathrm{i} \mathbf{G}_{(n,m)}\cdot \mathbf{r})}$, where $\mathbf{G}_{(n,m)} = n\,\mathbf{G}_1 + m\,\mathbf{G}_2\; (n, m \in \mathbb{Z})$ is a reciprocal lattice vector of the $1,064\,$nm 2D optical honeycomb lattice. The Schrödinger equation then takes the form
\be
    \bigg(\frac{1}{2m} \big(\mathbf{p} + \hbar\mathbf{q}\big)^2 + V(\mathbf{r}) \bigg) u_\mathbf{q}^n(\mathbf{r}) = E_\mathbf{q}^n u_\mathbf{q}^n(\mathbf{r}),
\ee
where $\mathbf{q}$ is the quasi-momentum vector, $n$ is the band index and $u_\mathbf{q}^n(\mathbf{r}) = \langle \br|u^n_\bq\rangle$ is the real space representation of the lattice periodic part $|u^n_\bq\rangle$  of a Bloch state $\ket{\psi_\mathbf{q}^n} = e^{\mathrm{i} \mathbf{q\cdot \hat r}} |u^n_\bq\rangle$. By expanding $V(\mathbf{r})$ and $u_\mathbf{q}^n(\mathbf{r})$ in the plane wave basis 
the Hamiltonian can be expressed explicitly as


\be
    H_{(n,m),(n',m')}(\mathbf{q}) = \bigg(\frac{2}{3} V_0 + \frac{\hbar^2 k^2}{2m} |\tilde{\mathbf{G}}_{(n,m)} + \tilde{\mathbf{q}}|^2 \bigg) \delta_{n,n'}\delta_{m,m'} -\frac{V_0}{9} \delta_{n,n'+j}\delta_{m,m'+l} \;\; \text{with } (j,l) \in \mathcal{J},
\ee
where $\mathcal{J} = \{(-1,0), (0,-1), (1,0), (0,1), (-1,-1), (1,1)\}$, the tilde implies that the quantity is normalized by the lattice laser wavenumber $k=2\pi/\lambda$ with $\lambda=1,064~\mathrm{nm}$ and $V_0$ as the optical lattice depth.  Here, $V_0<0$ because the lattice light is red-detuned with respect to the principal atomic resonances of $^{87}$Rb.

\begin{figure} 
    \centering
    \includegraphics[width=0.5\columnwidth]{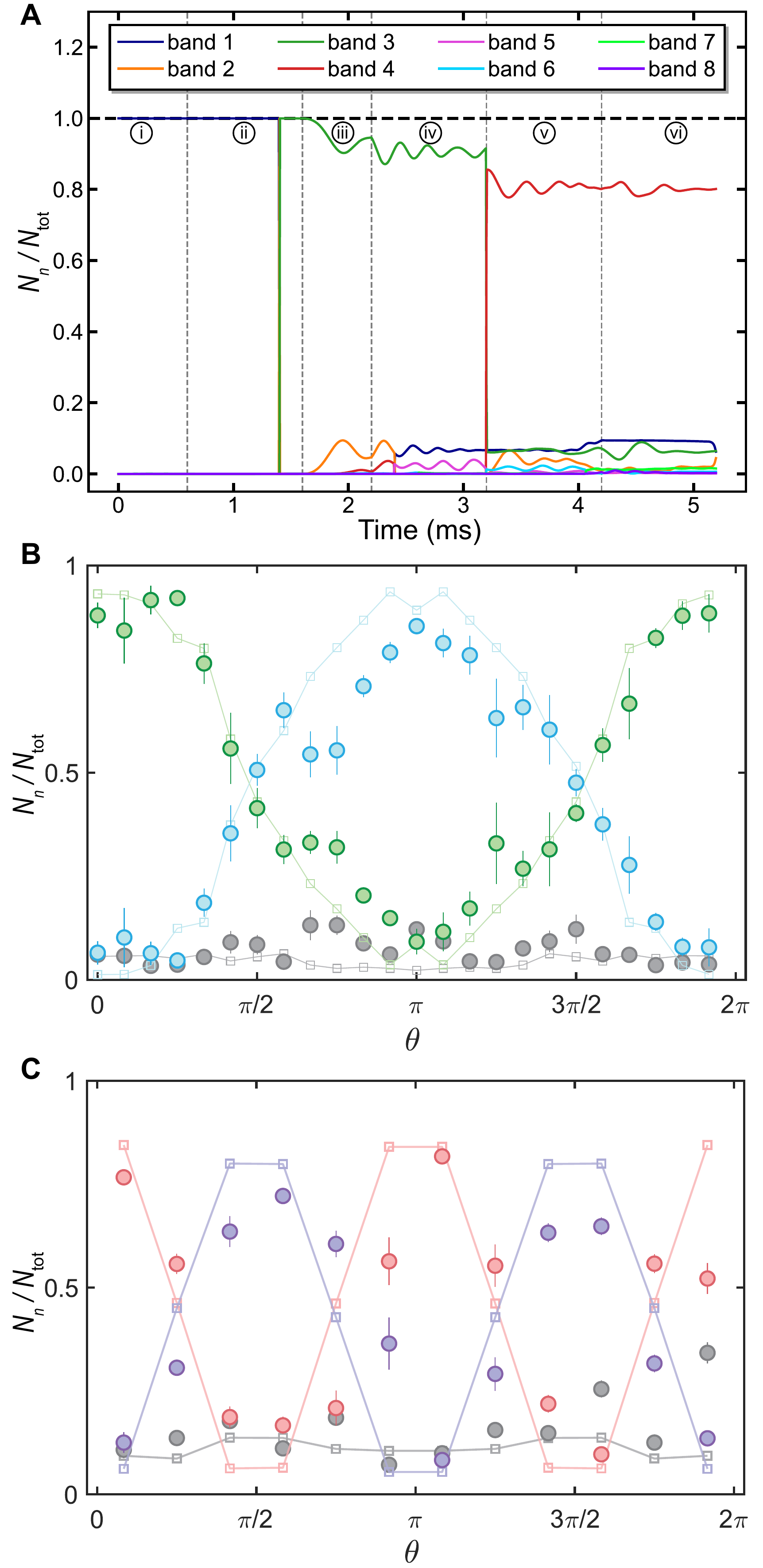}
    \caption{Simulation by decomposition of dynamic system state into Bloch eigenfunctions of $H(t)$. (\textbf{A}) Example of state evolution for $\theta = 255^{\circ}$ in Fig.~4 of the main text. The state $\ket{\psi(0)} = \ket{\mathbf{\Gamma},1}$ evolves over the following steps: i) lattice ramp down, ii) ramp to $\mathbf{q} = 1.25\,\mathbf{K}$, iii) rapid lattice intensity ramp up, iv) ramp to $\mathbf{q} = \mathbf{\Gamma}$, v) ramp to $\mathbf{q} = \mathbf{q}_{255\text{\textdegree}}^{r=0.9}$, before vi) band mapping. In v), the superscript labels the distance away from $\mathbf{\Gamma}$, in units of $\mathbf{q_{K}}$. Due to nonadiabatic evolution, bands other than $n=3,4$ are populated. (\textbf{B}) and (\textbf{C}) Simulation results (squares with dashed lines) for the experiments carried out in Fig.~2 and Fig.~4, respectively. Experiment data (circles) are copied here for reference.}
    \label{fig:Sim1}
\end{figure}

We simulate the system dynamics by truncating the plane wave basis such that it incorporates 81 different plane wave components and we divide the time evolution into time steps of $\tau = 5~\mu$s. We ensured that the simulation captures all relevant dynamics of the system with these settings.
The initial state of the system $\ket{\psi(0)}$ is often taken to be the ground band ($n=1$) eigenstate. We utilize spectral decomposition to speed up our computation. We apply $H(t)$ to the state such that $\ket{\psi(t)} = T\prod_{t'<t} e^{-\mathrm{i} H(t') \tau/\hbar} \ket{\psi(0)}$, where $T$ implies that the sum is properly time ordered.
With this method we can easily extract the populations of the state in each of the eigenstates $|u^n_\bq\rangle$ at each time step by first calculating the eigenstates and subsequently the overlap with the state $\ket{\psi(t)}$, as shown in Fig.~\ref{fig:Sim1}A. In that manner we can ensure that the effects we observe originate in the system dynamics and that we are able to quantify the effect of non-adiabatic population transfer. We use this technique to minimize excitations due to dynamics unrelated to the topological nature of the band structure singularities.

The simulated population transfer results for the experiments in Fig.~2 and Fig.~4 of the main text are plotted in Fig.~\ref{fig:Sim1}B and Fig.~\ref{fig:Sim1}C, with the experiment data reproduced in the same plot. Although these simulations do not involve any interaction effects, they nevertheless show no significant discrepancies to the observed data.

\section{Population dynamics and topology of the band touching points within the two-band models}
\label{section:2BandModel}
In this section, we show that all the main results presented in the main text can be qualitatively understood within an effective two-band model. To do that, we derive the equations which describe the band population of a particle moving in a lattice in the presence of an externally applied force within a two-band model. Relevant to our experiment, we only focus on the case of a constant force. In our derivation, we closely follow that from the supplementary materials from Ref.~\cite{li_bloch_2016}. Note that in the following sections $i$ is the imaginary number.

We start with the two-band lattice Hamiltonian of the form 

\be 
\hat H_0 = \sum_{n=1,2} \sum_{\bq} \ve_{\bq}^n |\psi_{\bq}^n \rangle \langle \psi_{\bq}^n|,
\ee 
where $|\psi_{\bq}^n \rangle$ are Bloch eigenstates of the $n$-th band at quasi-momentum $\bq$ having energy $\ve_{\bq}^n$. The Bloch states can be parameterized as $|\psi^n_\bq\rangle = e^{i \bq \cdot \hat\br} |u^n_\bq\rangle$, where $|u^n_\bq\rangle$ has periodicity of the lattice in real space and $\hat \br$ is the position operator. The Bloch states satisfy the normalization conditions $\langle \psi^n_\bq | \psi^m_{\bq'} \rangle = \delta(\bq - \bq')\delta_{nm}$ and $\langle u^n_\bq | u^m_\bq\rangle = \delta_{nm}$.

In the presence of an external constant force $\bF$ the Hamiltonian of the system reads as $\hat H = \hat H_0 - \bF \cdot \hat \br$, leading to the Schr{\"{o}}dinger equation

\be  
i\partial_t |\Psi(t)\rangle = (\hat H_0 - \bF \cdot \hat \br) | \Psi(t)\rangle,
\ee 
where we use the units with $\hbar = 1$.

We assume that at $t=0$ the particle has quasi-momentum $\bq_0$, implying that its wavefunction is given by

\be  
|\Psi(0)\rangle = \sum_{n=1,2} \varphi_n(0) |\psi_{\bq_0}^n\rangle,
\ee 
and the band population at $t=0$ equals $|\varphi_n(0)|^2$. In the presence of an external force, the quasi-momentum of the particle changes according to 

\be  
\bq(t) = \bq_0 + \bF t,
\ee 
meaning that the solution has form

\be  
|\Psi(t)\rangle = \sum_{n=1,2} \varphi_n(t) |\psi_{\bq(t)}^n\rangle.
\ee 
The Schr{\"{o}}dinger equation can then be rewritten as:

\be  
i\partial_t \left( \begin{array}{c} \varphi_1 \\ \varphi_2   \end{array}  \right) =   \left( \begin{array}{cc} \ve^1_{\bq(t)} - \xi^{11}_{\bq(t)}  & -\xi^{12}_{\bq(t)} \\ -\xi^{21}_{\bq(t)} &  \ve^2_{\bq(t)} - \xi^{22}_{\bq(t)}  \end{array}  \right)    \left( \begin{array}{c} \varphi_1 \\ \varphi_2   \end{array}  \right), \label{SMEq:Schrodingerequation}
\ee 
where we have defined 

\be  
\xi^{nm}_{\bq(t)} \equiv \bF \cdot \bA^{nm}_{\bq(t)} = i\langle u^n_{\bq(t)}|\partial_t| u^m_{\bq(t)}  \rangle, \label{SMEq:FdotA}
\ee 
and

\be  
 \bA^{nm}_{\bq} \equiv i\langle u^n_{\bq}|\partial_{\bq}| u^m_{\bq}  \rangle
\ee 
is the Berry connection. As we see from Eq.~\eqref{SMEq:Schrodingerequation}, it is precisely the off-diagonal components of the Berry connection that drives the population change of the two bands. 

\subsection{Linear band touching}

To study linear band touching, we apply the above analysis to the  two lowest bands of the typical band structure of the honeycomb lattice arising from the $s$-orbitals. Under the tight-binding approximation with only nearest neighbor hopping, the Hamiltonian of the (non-driven) system has the form 

\be  
H_0(\bq) = \left( \begin{array}{cc} 0 & M(\bq) \\ M^*(\bq) & 0    \end{array}  \right), \label{SMEq:H0swave}
\ee 
with

\be
M(\bq) = - t \left( 2 e^{iq_y a/2} \cos \frac{\sqrt{3} q_x a}2 + e^{- i q_y a}   \right),
\ee 
where $-t$ is the hopping integral and $a$ is the lattice constant.

\begin{figure*}
  \centering
  \includegraphics[width=\columnwidth]{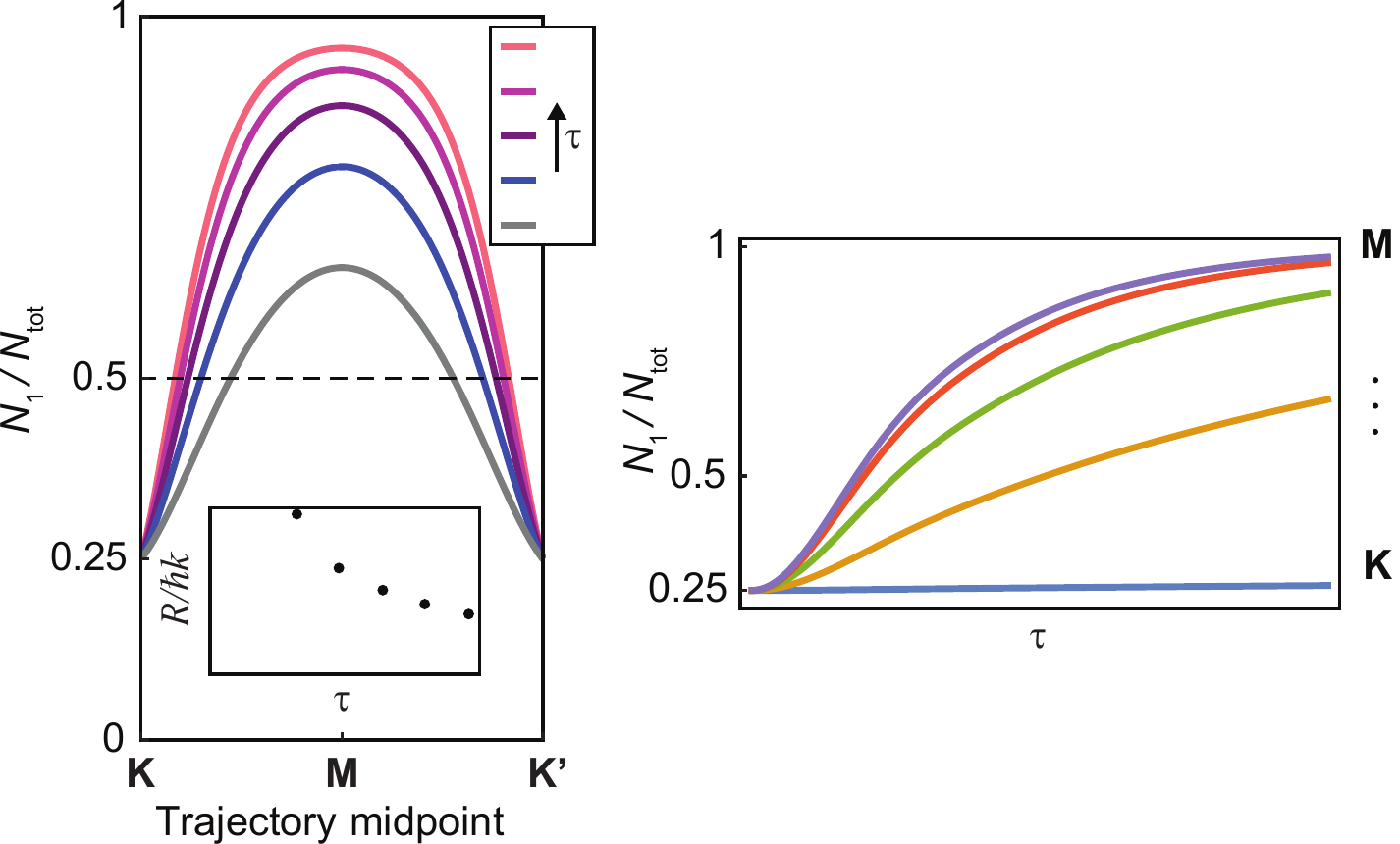}
    \caption{Final band population as a function of the acceleration time $\tau$ and different trajectories shown in Fig.~\ref{fig:all_deltas}A calculated within a two-band model, Eq.~\eqref{SMEq:H0swave}.
    Left: Population of the $n=1$ band as a function of the trajectory midpoint at different fixed $\tau$ and the size of the singularity $R$ as a function of $\tau$. The black arrow indicates the increase of $\tau$. The results are in good qualitative agreement with the measured data shown in Figs.~3B,C of the main text. Right: The normalized $n=1$ band population presented as a function of the acceleration time for different fixed trajectories. Again, the result of the calculation is in good qualitative agreement  with the experimental data shown in Fig.~\ref{fig:all_deltas}B.
    } 
  \label{FigSM:contrastsize}
\end{figure*}

With this model Hamiltonian, we use Eq.~\eqref{SMEq:Schrodingerequation} to calculate the final band population as the particles move along the trajectories shown in Fig.~\ref{fig:all_deltas}A (or Fig.~3A of the main text), assuming that originally they were prepared in the $n=1$ state. The result as a function of the trajectory midpoint and the acceleration time $\tau$ is presented in Fig.~\ref{FigSM:contrastsize} and is in good qualitative agreement with Figs.~3B,C of the main text and Fig.~\ref{fig:all_deltas}B. The only difference appears at short acceleration times, which is quite natural since at such large acceleration the population of the higher bands becomes substantial and must be taken into account.

The Dirac singularities are located at the points $\mathbf{q}_{\mathbf{K'}} = \left( 4\pi/3\sqrt{3}a, 0  \right)$ and $\mathbf{q}_{\mathbf{K}} = \left(2\pi/3\sqrt{3}a, 2\pi/3a \right)$ in the Brillouin zone and those related by the reciprocal lattice vectors. Near the $\mathbf{K}$-point, the Hamiltonian can be expanded to linear order in quasi-momentum $\bk \equiv \bq - \mathbf{q}_{\mathbf{K}}$ as

\be  
H_0(\bk) \approx v_g\left( \begin{array}{cc} 0 & k_+ \\ k_- & 0    \end{array}  \right), \qquad k_{\pm} = k_x \pm i k_y, \label{SMEq:H0Dirac}
\ee 
where $v_g =3at/2$ is the group velocity at the Dirac point. The wave functions of the $n=1$ and $n=2$ bands are given by

\be  
|u_\bk^1\rangle = \frac1{\sqrt{2}} \left( \begin{array}{c} -k_+/k \\ 1 \end{array}\right), \qquad |u_\bk^2\rangle = \frac1{\sqrt{2}} \left( \begin{array}{c}  k_+/k \\ 1 \end{array}\right), \label{SMEq:uDirac}
\ee 
with $k_{\pm} \equiv k_x \pm i k_y,$ $k \equiv |\bk| = \sqrt{k_x^2 + k_y^2}$, and the components of the Berry connection equal

\be  
\bA_{\bk}^{11} = \bA_{\bk}^{22} = - \bA_{\bk}^{12} = - \bA_{\bk}^{21} = 
\frac1{2(k_x^2 + k_y^2)}\left( \begin{array}{c} k_y \\ -k_x   \end{array}  \right).
\ee 
We emphasize that while the specific form of the wave functions and Berry connection depend on the gauge of the wave functions, the final answer for the observables, e.g. band population, is gauge-independent.

When a particle moves exactly towards or away from the singularity in momentum space, it does not experience any interband transitions away from the singular point itself since $\bF \cdot \bA_{\bk}^{mn} =0$. This is a general feature of any  homogeneous Hamiltonian satisfying the relation $H_0(s \bk) = s^\alpha H_0(\bk)$. Indeed, if choosing a proper gauge like the one in Eq.~\eqref{SMEq:uDirac}, the eigenstates of the Hamiltonian depend only on the direction of the quasi-momentum, $\hat \bk = \bk /|\bk|$, but not on its absolute value. Then, according to Eq.~\eqref{SMEq:FdotA},  $\bF \cdot \bA_{\bk}^{mn} =0$ because $\hat{\bk}(t)$ and, consequently, $|u_{\bk(t)}^m\rangle$ do not depend on $t$. This implies that all the transitions happen exactly at the singularity. This conclusion is true provided we stay in the region of the Brillouin zone where the Hamiltonian of the system can be approximated by a homogeneous function (i.e., we must remain sufficiently close to the singular point). Assuming that the particle moving along the direction $\hat \bk$ scatters off the singularity in the direction $\hat \bk'$ at time $t_0$, the transition probability between the two bands can then be found from the continuity condition $|\Psi(t_0-0)\rangle = |\Psi(t_0+0)\rangle$, which translates into                            
\begin{align}
&\varphi_1(t_0-0) |u_\bk^1\rangle + \varphi_2(t_0-0) |u_\bk^2\rangle \nonumber \\ = &\varphi_1(t_0+0) |u_{\bk'}^1\rangle + \varphi_2(t_0+0) |u_{\bk'}^2\rangle.
\end{align}
If the particle was originally in the $n=1$ band, $\vp_2(t_0-0) = 0$, then the transition probability is given by

\be  
d^2(\bk, \bk') = |\vp_2(t_0+0)|^2 = |\langle u_{\bk'}^2 |   u_{\bk}^1  \rangle|^2 = 1 - |\langle u_{\bk'}^1 |   u_{\bk}^1  \rangle|^2.
\ee 
This quantity is exactly the Hilbert-Schmidt quantum distance between the states $| u_{\bk'}^1  \rangle$ and $| u_{\bk}^1  \rangle$~\cite{provost_riemannian_1980,hwang_wave-function_2021}. For two infinitesimally close states $\bk$ and $\bk' = \bk + \mathrm{d}\bk$, it reveals the gauge-invariant quantum metric tensor $g_{ij}(\bk)$:

\be  
d^2(\bk, \bk+ d \bk) = g_{ij}(\bk) \mathrm{d}k_i \mathrm{d}k_j,
\ee  
which is defined as 
\be 
g_{ij}(\bk) = \frac12[\langle \partial_i u_\bk | \partial_j u_\bk  \rangle -  \langle \partial_i u_\bk | u_\bk \rangle \langle u_\bk | \partial_j u_\bk  \rangle + (i \leftrightarrow j) ].
\ee

In the case of the isotropic Dirac dispersion, the quantum distance is expressed through  the angle $\theta$ between the  momenta $\bk$ and $\bk'$:

\be  
d^2_{\text{Dirac}}(\bk, \bk') = \sin^2 \frac{\theta}2.
\ee 
This result is in perfect agreement with the pseudo-magnetic field picture discussed in the main text and leads to the same prediction.

It is worth noting that such a clear and simple relation between the band population change and the quantum distance (and Hilbert space geometry) is, strictly speaking, only possible in the vicinity of the singular point, where the Hamiltonian can be approximated by a homogeneous function. In the experiment, on the other hand, the start and end points of the trajectory in the Brillouin zone clearly lie outside the singular region. Nevertheless, all the conclusions derived above remain valid provided the particle's motion is sufficiently slow. Indeed, outside the singular region, the system is effectively gapped. Hence, according to the adiabatic theorem, the interband transitions are rare if the drive is slow.

\subsection{Quadratic band touching}

The entire analysis for the quadratic band touching point is similar to that of the previous section, yet the results show some important differences, which we highlight and discuss in detail below. 

The lowest-energy quadratic band touching is formed by the $n=3$ and $n=4$ bands at the $\mathbf{\Gamma}$-point. To derive the effective low-energy theory for the QBTP, we follow Refs.~\cite{zhang_honeycomb_2014,sun_topological_2009} and start with the tight-binding model band structure that arises from the $p_x$ and $p_y$ orbitals. We only include the $\sigma$-bonding and nearest-neighbor hopping for simplicity. The Hamiltonian in this case takes the form

\be
H = t_{\sigma}\sum_{i=1}^3\sum_{\vec{r} \in A} \left\{ p^{\dagger}_{i} (\vec{r}) p_{i} (\vec{r} + a \hat{e}_i) + \text{H. c.}\right\},
\ee
where $p_i = (p_x \hat{e}_x + p_y \hat{e}_y) \cdot \hat{e}_i$ are 
the projections of the $p$ orbitals parallel to the nearest-neighbor bond directions $\hat e_i$, $p_{x,y}^\dagger$ represent the creation operators of atoms on the corresponding orbitals, and $t_\sigma$ is the $\sigma$-bonding strength. Vectors $\hat{e}_{i}$ are three unit vectors from one $A$ site to its three neighboring $B$ sites and are given by $\hat{e}_{1,2} = \pm \frac{\sqrt{3}}{2} \hat{e}_x + \frac{1}{2} \hat{e}_y $ and $\hat{e}_3 = - \hat{e}_y$.


The Fourier transform of the Hamiltonian can be written as

\be
H(\bk) = \begin{pmatrix} 0 &  h(\bk)\\
h^{\dagger}(\bk) & 0\end{pmatrix}, h(\bk) =\frac{t_{\sigma}}{2} \sum_{i=1}^3 e^{i a \bk \cdot \hat{e}_i} \left( I + \sigma_x \cos 2 \theta_i + \sigma_y \sin 2 \theta_i  \right), \label{SMEq:Hquadfull}
\ee
where $\sigma_{x,y}$ are Pauli matrices in the chiral orbital basis $(p_-/\sqrt{2},p_+/\sqrt{2})$, $p_{\mp} = p_x \mp i p_y$, $I$ is the identity matrix, and we have defined $\theta_{1} = \pi/6$, $\theta_{2} = 5\pi/6$, and $\theta_3 = 3\pi/2$. Importantly, this Hamiltonian is symmetric under time-reversal, inversion, and six-fold rotations.

The Hamiltonian in Eq.~\eqref{SMEq:Hquadfull} exhibits a quadratic band touching at $\bk = 0$. Expanding to $\mathcal{O}(k^2)$ we find

\be
    h(\bk) = \frac{3 t_{\sigma}}2 \begin{pmatrix}  1 & 0 \\ 0 & 1 \end{pmatrix} + \frac{3  t_{\sigma} a}4 \begin{pmatrix}  0 & k_+ \\ -k_- & 0 \end{pmatrix} - \frac{3 t_{\sigma} a^2 }{16} \begin{pmatrix}  2 k^2 &  k_-^2 \\ k_+^2 & 2 k^2 \end{pmatrix} +  \mathcal{O}(k^3).
\ee
After performing a unitary rotation in the sublattice space (two different sublattices originate from non-equivalent sites $A$ and $B$ in the primitive cell of the honeycomb lattice), we obtain

\begin{align}
&H(\bk) \to U^\dagger H(\bk) U \approx \frac{3 t_\sigma}2 \begin{pmatrix} 1 - \frac{(a k)^2}4 &   - \frac{(a k_-)^2}8 & 0 & \frac{a k_+}2   \\ - \frac{(a k_+)^2}8 & 1 - \frac{(a k)^2}4 &   -\frac{a k_-}2 & 0 \\ 0 & -\frac{a k_+}2 & - 1 + \frac{(a k)^2}4 &  \frac{(a k_-)^2}8 \\  \frac{a k_-}2 & 0 & \frac{(a k_+)^2}8 & - 1 + \frac{(a k)^2}4  \end{pmatrix}, \nonumber
\end{align}
where $U = (I-i \tau_y)/\sqrt{2}$ and $\tau_y$ is the Pauli matrix in the sublattice space. Rewritten in this form, the Hamiltonian is block-diagonal: the lower right (upper left) block describes quadratic band touching at energy $- 3 t_\sigma/2$ ($3 t_\sigma/2$), while the off-diagonal blocks describe virtual hopping between the low- and high-energy bands. Accounting for such hopping within second-order perturbation theory~\cite{mccann_landau-level_2006,hwang_wave-function_2021}, we end up with the effective low-energy quadratic Hamiltonian describing just two lower bands (which correspond to $n=3$ and $n=4$) that exhibit a QBTP:

\be  
H_{\text{eff}}(\bk) \approx \frac1{4m} \left( \begin{array}{cc} k^2 &  k_-^2 \\ k_+^2 & k^2  \end{array}   \right), 
\ee 
where the effective mass is given by $m^{-1}= 3 t_{\sigma} a^2/4$ and the constant term $-(3 t_\sigma/2) I$ has been dropped. The eigenenergies of this Hamiltonian are given by $\ve_3 = 0$ and $\ve_4 = k^2/2m$, i.e., we have a band touching between a quadratically-dispersing band and a flat band. The eigenvectors equal

\be  
|u_\bk^3\rangle = \frac1{\sqrt{2}} \left( \begin{array}{c} - k_-^2/k^2 \\ 1 \end{array}\right), \qquad |u_\bk^4\rangle = \frac1{\sqrt{2}} \left( \begin{array}{c} k_-^2/k^2 \\ 1 \end{array}\right), \label{SMEq:uquad}
\ee 
resulting in the Berry connection of the form

\be  
\bA_{\bk}^{33} = \bA_{\bk}^{44} = - \bA_{\bk}^{34} = - \bA_{\bk}^{43} = 
\frac1{k_x^2 + k_y^2}\left( \begin{array}{c} - k_y \\ k_x   \end{array}  \right).
\ee 
The Berry phase around the QBTP is twice that around the LBTP, i.e., equals $\pm 2\pi$. Since the Berry phase is defined only mod $2\pi$, such a characteristic is not particularly useful in distinguishing a topologically-trivial QBTP from a non-trivial one. Instead, as we discuss below, one can use a gauge-independent integer winding number, provided certain symmetries are present. 

Repeating the same derivation as for the LBTP, we find that the population of the $n=4$ band after scattering off the singularity (assuming the particles are originally in the $n=3$ band) and the quantum distance are given by 

\be  
d^2_{\text{quad}}(\bk, \bk') = |\langle u_{\bk'}^3 |   u_{\bk}^4  \rangle|^2 = \sin^2 \theta.
\ee 

It is worth discussing now certain important differences between the geometries of the linear and quadratic band crossings. For a comprehensive analysis of this and related questions we refer to Ref.~\cite{hwang_wave-function_2021}, while we only focus on the most important points. 

In general, linear band touchings are always characterized by a Berry phase of $\pi$ (mod $2\pi$) around them, and the maximal quantum distance between the states encircling the point always equals $d^{\max}_{\text{Dirac}} = 1$, meaning that there are always orthogonal states no matter how close we are to the singularity. The linear singularities are always topological in this sense and no symmetries play any role in this conclusion. 

Quadratic band touching points are quite different in many ways. Despite the common belief, the Berry phase around the QBTP, in general, does not have to be quantized to integers of $2\pi$. Related to this fact, the maximal quantum distance between the states near the band touching point, in general, can take an arbitrary value between 0 and 1, and so can the amplitude of the transition probability oscillations $|\langle u_{\bk'}^3 |   u_{\bk}^4  \rangle|^2$.

However, this arbitrariness is removed when certain symmetries are present in the system. In particular, when time-reversal and rotational $C_6$ symmetries are present, which is the case in our system, the Berry phase becomes quantized in units of $2\pi$, and the quantum distance can only take values 0 or 1. We emphasize that the quantum distance is a more useful quantity for characterizing the non-trivial topology of the band touching. Indeed, while the Berry phase is defined only mod $2\pi$, its values of $\pm 2\pi$ can be easily changed to 0 by a simple gauge transformation. It implies that the Berry phase cannot reliably distinguish between a trivial QBTP with $d_{\max}=0$ and a non-trivial one with $d_{\max}=1$. 

Equivalently, when time-reversal and rotational $C_6$ symmetries are present, the non-trivial topology of the QBTP can be characterized by a well-defined gauge-independent winding number~\cite{hwang_wave-function_2021}. In general, the effective two-band Hamiltonian can always be rewritten as a pseudo-spin in a pseudo-magnetic field $\bB(\bk)$, $H_{\text{eff}}(\bk) = B_0(\bk) \cdot I + \bB(\bk) \cdot {\boldsymbol \sigma}$, where ${\boldsymbol \sigma}$ is the vector of Pauli matrices. If the aforementioned symmetries are present, one can choose a basis such that $\left\{ \bB(\bk) \cdot {\boldsymbol \sigma}, \sigma_z \right\} = 0$, i.e., $B_z(\bk) = 0$. In this case, the pseudo-magnetic field lies entirely in the transverse pseudo-spin plane, where it traces out a great circle on the unit Bloch sphere as $\bk$ goes around the band touching point. In this sense, the winding of $\bB(\bk)$ around the origin (singularity) is a topological property since it cannot be changed smoothly. The winding number counts how many times $\bB(\bk)$ encircles the origin and is well defined as long as $\bB(\bk)  \neq 0$ on a trajectory that encompasses the singularity. It can be explicitly defined as~\cite{chiu_classification_2014,chiu_classification_2016,montambaux_winding_2018}

\be  
w = \frac{i}{2\pi} \oint d\bk  \nabla_\bk \ln B_-(\bk),
\ee
where the contour of integration encircles the singularity and $B_- = B_x - i B_y$. Near the QBTP in our system, $B_0(\bk) = k^2/4m$,  $\bB(\bk)= (1/4m) (k_x^2 - k_y^2, 2 k_x k_y,   0)^T$, and $B_-(\bk) = k_-^2/4m$, and so the winding number for the QBTP $w_{\text{quad}} = 2$. For comparison, the winding number for Dirac Hamiltonian~\eqref{SMEq:H0Dirac} equals $w_{\text{Dirac}} = -1$.

The amplitude of the oscillations measured experimentally and shown in Fig.~4C of the main text differs from 1 due to nonadiabaticity in the lattice acceleration steps and, particularly, in the band mapping procedure. Nevertheless, according to the discussion above, the
non-zero amplitude and two cycles of oscillations combined with our lattice symmetries unambiguously indicate that the topological winding number around the QBTP we study is well-defined and equals 2.



\newpage


\end{document}